\documentclass[12pt]{article}

\def\noi{\noindent}

\def\skip{\vspace{2mm}}
\newcount\u
\def\Remark{\skip\noi{{\bf{Remark \number\u.}}} \advance\u by 1}

\usepackage{amsmath,amssymb,amscd}

\usepackage{graphics}                   
\usepackage[dvips]{epsfig}
                   
\usepackage{delarray,hhline,latexsym}

\usepackage{float}

\usepackage{enumerate}

\usepackage{fancyhdr}

\def\noi{\noindent}

\newtheorem{props}{Proposition}

\newcounter{defnctr}
\newtheorem{defns}[defnctr]{Definition}

\newtheorem{corrs}[props]{Corollary}

\newtheorem{lemms}[props]{Lemma}

\newcounter{exctr}
\newtheorem{es}[exctr]{Example}

\newtheorem{ess}[exctr]{Examples}

\begin{document}

\title{A Bayesian Game without $\epsilon$-equilibria}
\author {Robert Samuel Simon and Grzegorz Tomkowicz}

\maketitle
\thispagestyle{empty}
\vfill

\noi  Robert Simon 
\newline London School of Economics
\newline 
Department of Mathematics\newline 
Houghton Street\newline 
 London WC2A 2AE\newline                    
email: R.S.Simon@lse.ac.uk
\vskip.5cm

 \noi Grzegorz Tomkowicz\\
  Centrum Edukacji $G^2$\\
        ul.Moniuszki 9\\
        41-902 Bytom\\
        Poland\\
      e-mail: gtomko@vp.pl

\vfill

\date{}

\setcounter{page}{-1}
  \noi

\newpage
\vskip2cm
\thispagestyle{empty}
\begin{center}
{\bf Abstract}
\end{center}
\vskip1cm

\noi  We present a three player Bayesian game for which there is no 
 $\epsilon$-equilibria in Borel measurable strategies for small enough 
 positive $\epsilon$, however 
 there are non-measurable equilibria.

\vskip2cm

\noi {\bf Key words}: Bayesian games, non-amenable  semi-group action,
  equilibrium existence

   \newpage
\section {Introduction}
 Game theory is about strategies and the expected payoffs resulting 
 from these strategies. The fundamental concept of game theory 
 is that of an equilibrium, strategies for each player such that 
 no player prefers to switch to another strategy, given that 
 the strategies of the other players remain fixed. An $\epsilon$-equilibrium 
 for any $\epsilon \geq 0$ is defined in the same way --   
 no player prefers by more than $\epsilon$ to switch to another 
 strategy. An equilibrium is a $0$-equilibrium. 
\vskip.2cm 
 \noi 
What do we 
 mean  by  a player preferring   some strategy over  another by 
 at least $\epsilon$? 
 Implied by a preference  is an evaluation in an ordered 
field, usually  
   the real numbers. Conventionally we assume that there 
 is a topological probability space on which the game takes place, that each 
 player is assigned a Borel sigma field corresponding to its  
 knowledge of the space, and that    
 that the players choose strategies that are measurable with respect 
 to  their respective Borel sigma fields. Given fixed measurable 
 strategies of the other players, the evaluation of 
 a strategy for a player is  global,  an integration of a function 
 over the whole probability space.  When the players 
 attain an $\epsilon$-equilibrium by this criteria,    
 a {\em global} or {\em Harsanyi}
 $\epsilon$-equilibrium is obtained.   \vskip.2cm 
\noi On the other hand, a player could orient itself locally to some minimal 
 sets in its sigma algebra, known as {\em information sets} 
 (should they exist), and maximise its payoff 
 with regard to  these sets and its knowledge of them (which we 
 presume is consistent with its knowledge of the whole probability space 
 through regular conditional distributions). When each player at each 
 such minimal set cannot obtain more than $\epsilon$ with the payoff 
  evaluated locally at  
 this set, a Bayesian $\epsilon$-equilibrium is obtained. 
   \vskip.2cm
\noi Bayesian games are ancient, most card games being good examples. 
 J. Harsanyi (1967) introduced a global theoretical 
  perspective to  Bayesian games, 
 the origin of the term Harsanyi equilibrium. P. Milgrom and R. Weber (1985) 
 asked implicitly
 the question whether Bayesian games always have measurable 
 equilibria after proving  existence for a
 special class of Bayesian games and analysing  a game which 
 did not belong to that class.
 R. Simon (2003) demonstrated an example of a three player 
  Bayesian game for which 
 there is no Borel measurable equilibrium however there are many 
 non-measurable equilibria. This example was made possible by a structure 
  of knowledge for which always the players  have common knowledge of 
 a countable set of measure zero. These sets of measure zero give no 
 orientation for compiling  the many  local equilibria toward Borel 
    measurable 
 strategies. Indeed, elementary applications 
 of ergodic theory showed that 
Borel  measurable equilibria could not exist.  
\vskip.2cm 
\noi A significant advance  was performed by Z. Hellman (2014). He 
 showed that there  is a two-player Bayesian game with Bayesian equilibria 
 but  no 
  Bayesian $\epsilon$-equilibrium  that is also Borel measurable for small 
 enough positive $\epsilon$. 
   This discovery 
 was  advanced further by   Z. Hellman and 
 J. Levy (2016), in which it was demonstrated  that a broad class of 
  knowledge structures support games for which the same holds. This 
 paper  serves well as a general source 
 to the structure, problems, and history of  Bayesian games, 
 especially in their relation to countable equivalence relations that
 are amenable.   
\vskip.2cm 
\noi It is required of a 
 Bayesian $\epsilon$-equilibrium that  throughout the space each player  
 cannot gain locally more than $\epsilon$ through an alternative strategy. 
 However in a Harsanyi $\epsilon$ -equilibrium there could be  
  a player who, according to the strategies defining the 
 $\epsilon$ equilibrium, could improve its payoff by as much as 
  $B>0$ at a set of measure 
 no more than $\epsilon/B$. A Borel measurable Bayesian $\epsilon$-equilibrium 
 is a Harsanyi $\epsilon$-equilbrium, but not vice versa. 
 Indeed it is not difficult to 
 show that for every $\epsilon>0$ there are Harsanyi $\epsilon$-equilibria 
 to the Hellman game cited above.  \vskip.2cm 
\noi          One can perceive  sets of very small measure  
 where a player can act foolishly as a kind of firewall, absorbing 
 the problems of the measurability requirement. Amenable 
 structures tend to allow for such
 firewalls; for example with the closely related topic of 
  Borel colouring; see   Kechris, et all (1999). 
Therefore we would not have expected to find a game  example lacking  
  Harsanyi $\epsilon$  equilibria (yet possessing Bayesian equilibria) 
 without utilizing  a non-amenable structure.   
\vskip.2cm

\noi In this paper we demonstrate that there is a Bayesian game 
 played on a  topological probability space  $\Omega$ for which there are 
 no Harsanyi $\epsilon$-equilibria for all $\epsilon \leq \frac 1 {1000}$, and 
 yet  there are non-measurable 
 Bayesian equilibria that employ  pure 
 strategies almost everywhere (pure meaning all weight given to one action). 
 Because $\Omega$ is a Cantor set and the concern is  
 the existence of  approximate equilibria, 
 moving from Borel to Lesbegue measurability does not alter  
 the result. 
\vskip.2cm

\noi As long as a game has an Harsanyi $\epsilon$-equilibrium for 
 every positive $\epsilon$ there is an equilibrium payoff, namely 
a cluster point of payoffs corresponding to the $\epsilon$-equilibria as 
 $\epsilon$ goes to $0$. 
By this interpretation of an equilibrium payoff, 
 ours is a Bayesian game that has equilibria, 
 but no equilibrium payoff. \vskip.2cm 
 
\noi With our example, there is  no proper subset of the probability 
 space for which  
 the players  have  common knowledge,    
 hence the  arguments used  
 are   different from that of previous Bayesian games that lack  
 Harsanyi equilibria but have Bayesian equilibria (which do utilize 
 countable equivalence relations).
  Nevertheless we use a non-amenable semi-group action 
 that contains some similarities to the structure of the Hellman example. 
\vskip.2cm 
\noi The rest of this paper is organised as follows. 
 In the next section we define the game, followed in the third section by 
 a proof that this game has no Harsanyi $\frac 1 {1000}$-equilibrium. 
In the fourth section we show that 
 it does have non-measurable  Bayesian equilibria. The concluding 
 fifth section is a presentation of open problems. 
   
\section { The Game:} \vskip.2cm  
\noindent Let $G^+$ be the free semi-group generated by the non-negative  
 powers of the two generators $T_1$ and $T_2$, with $e$ the identity 
 included in $G^+$.  Let $X$ be 
 the space  $\{ 0,1\}^{G^+}$, with 
 an $x\in X$ a collection of the form  
 $(x^e, x^{T_1} , x^{T_2}, x^{T_1 T_2}, 
 x^{T_2 T_1},$ \newline $ x^{T_1^2} , x^{T_2^2},   \dots )$
 with $x^U \in \{ 0,1\}$ for every 
 $U\in G^+$. 
 For both  $i=1,2$ define 
  $T_i:X \rightarrow X$ to be   
 the   shift: $T_i(x)^V = x^{T_iV}$ for all $V\in G^+$, and 
 for every $U,V\in G^+$  define $U : X \rightarrow X$ by 
 $U(x)^V = x^{UV}$. This defines a right action of $G^+$ on $X$, meaning 
 that $UV(x) = V \circ U (x)$. For every $x\in X$ define $G^+(x)$ to be 
 the countable set 
$\{ U(x)\ | \ U\in G^+\}$.  Very important to the structure of $X$ is 
 that for every $x,y\in X$ there are two $z_1, z_2 \in X$ 
 such that $T_1(z_i) =x$ and $T_2 (z_i) =y$ for both $i=1,2$. We call the points
 \emph{the twins determined} by $x$ and $y$.   We place 
 the canonical probability distribution $m$ on $X$  
which gives $\frac 12$ to each $0$ or $1$ placed in each position 
 of $G^+$ and independently, so that a cylinder defined 
 by $n$ positions is given the probability $2^{-n}$. 
 With this probability distribution, we 
 see that all $U\in G^+$ are measure preserving actions, 
 meaning that $m(U^{-1} A) = m(A)$ for all Borel subsets $A$.     
   \vskip.2cm 
\noi Of special importance,  the 
 indendence probability assumption on each  position   implies that 
 the  distribution $m$  on $X$ can be reconstructed 
    from  the  measure preserving property of  $T_1:X \rightarrow X$ and 
 $T_2:X \rightarrow X$,  its product distribution $m^2$, 
 combined with  
 the equal probability given to both twins. \vskip.2cm    

\noi Let $D$ be the set   
$D:= \{ r, g\}$, $r$ for red $r$ and $g$ for green. 
   The probability space on which the game is played  will be
 $\Omega := D\times  X$. We define the topology 
 on $\Omega$ to be that induced by the clopen 
 (closed and open) sets defined by the set $D$  
 and the cylinder sets of $X$, so that $\Omega$ is a Cantor set.  
 We define the canonical  probability distribution $\mu$ on 
 $\Omega$ so that   
for each choice of $d\in D$ and $0$ or $1$  in 
  $n$ distinct positions the  probability for this 
 cylinder set  will be $\frac 12\  2^{-n}$.  
  For example, $\mu$ gives 
 the set $\{ (r, x) \ | \ x^e=0, x^{T_1}=1 \}$ the  probability 
 $\frac 1 {8}$. The measure 
 $\mu$ is the {\em common prior} for the game, meaning the Borel 
 probability measure by which the game is defined.  
 
  \vskip.2cm 

 \noi There are three  players, labelled   $G_0, R_1, R_2$.   
 The information sets of each player are defined as   follows. 
   For each $x\in X$ Player $G_0$ considers  $(g, x)$ and 
 $(r,x)$ 
 possible, and with equal $\frac 12$  probability, and
 these two  points constitute its information set. For each $i=1,2$ and 
 each $x$ Player $R_i$ consider  $(r, x)$  and $\{ g\} \times 
  T_i^{-1} (x)$ 
 possible,  with  the point     $(r, x)$ 
  and the set $\{ g\} \times 
  T_i^{-1} x$ of equal $\frac 12$ probability,
 and this pairing of a point  with the corresponding 
  Cantor set is its information set. 
 Notice that this belief by the player $R_i$ 
  is   consistent with 
 the probability distribution on $\Omega$, as the measure preserving 
 property of the $T_i$ implies that $m(T_i^{-1} (A)) = m(A) $ for all Borel 
 subsets $A$ of $X$.   
  If $B$ is the information set of a player, it means 
 that this player cannot distinguish between any two points of this set and 
 therefore has to conduct the same behaviour throughout the set. The Borel 
 sigma algebra defining the knowledge of a player is that induced by 
 these information sets, meaning the collection of all Borel sets $B$ 
 such that every information set of this player 
  is either inside of $B$ or disjoint 
 from $B$.  
 \vskip.2cm 
\noi All players have only two actions. The 
  red players $R_1$ and $R_2$  have  the   actions ${\bf a_0}, {\bf  a_1}$
 and the 
 green player $G_0$  has the  actions ${\bf b_0}$, ${\bf b_1}$.
   \vskip.2cm 
\noindent 
  For either   player $R_i$ the only     payoff  that matters is that 
  obtained  
 at those states  
labelled  $r$, and for  
  the  player $G_0$ the same is true for those   states  
  labelled $g$. There are two equivalent approaches to be taken, 
 illustrated for a player $R_i$.  
 Either  the payoff obtained at $(r, x)$, described below, 
  is duplicated  at all the 
other points in the same information set, namely 
 the set $\{ g\} \times  T^{-1}_i (x)$, or 
 the payoffs obtained at $(r,x)$, described below, is multiplied by 
 $2$ and at all  other points in the same information set the payoff 
 is $0$. Though the latter interpretation may be better suited to 
 some theoretical approaches, as it employs the probability 
 distribution $\mu$ on $\Omega$, we 
  will assume throughout the former equivalent interpretation 
(and for Player $G_0$ as well). This will 
 allow us to focus on the set $X$ and its probability distribution $m$.   
   \vskip.2cm 

\noi 
 The ${\bf a_0}$ and ${\bf a_1}$  pertain to   
 actions of  Player $G_0$ at both $(g, x)$ and $(r, x)$. If 
  $x^e=0$ then the payoff matrices 
 for  the players  $R_i$ at the  states $(r, x)$ are 
 
 $$R_1 \quad \quad \begin {matrix} & {\bf a_0}  & {\bf a_1}  \cr 
{\bf b_0} & 300  & 0  \cr 
 {\bf b_1} & 0 & 100  \end{matrix} \quad \quad \mbox{ and } \quad \quad   
R_2  \quad \quad 
 \begin {matrix} & {\bf a_0}  & {\bf a_1}  \cr 
{\bf b_0} & 100  & 0  \cr 
 {\bf b_1} & 0 & 300  \end{matrix}.$$ 
If $x^e=1$ then the payoff matrices at $(r,x)$ 
 are reversed: 
 $$R_1  \quad \quad \begin
 {matrix} & {\bf a_0}  & {\bf a_1}  \cr 
{\bf b_0} & 100  & 0  \cr 
 {\bf b_1} & 0 & 300  \end{matrix} \quad \quad  \mbox{ and } \quad \quad   
R_2  \quad \quad 
 \begin {matrix} & {\bf a_0}  & {\bf a_1}  \cr 
{\bf b_0} & 300  & 0  \cr 
 {\bf b_1} & 0 & 100  \end{matrix} .$$

\vskip.2cm 
\noi   
 \noindent More  complex are the  payoffs of the player $G_0$ at a 
 state labelled  $g$.
   The matrix is three dimensional,
 meaning that it is a $2 \times 2 \times 2$ matrix. 
 We need only to  describe a 
   $2\times 2$ matrix corresponding to each action of 
 the $G_0$ player. The rows and columns stand for the actions of 
 the $R_1$ and $R_2$ players, respectively. Those  actions 
 ${\bf a_0}$ and ${\bf a_1}$ are performed by 
 the $R_1$ player at both  $(g,x)$ and $(r, T_1 x)$ and by 
 the $R_2$ player at both $(g,x)$ and $(r, T_2 x)$. First we describe 
 the payoff matrices  if $x^e = 0$:  
    
$$  {\bf b_0} \quad \quad \begin {matrix} & {\bf a_0}  & {\bf a_{1} }  \cr 
{\bf a_0} & 1000 & 0 \cr 
 {\bf a_{1} } & 0 & 2000   \end{matrix} \quad \quad \quad \quad {\bf b_1} 
 \quad \quad \begin {matrix} & {\bf a_0}  & {\bf a_{1} }  \cr 
{\bf a_0} & 0 & 1000 \cr 
 {\bf a_{1} }& 2000 & 0   \end{matrix}$$
\vskip.2cm 
 \noi 
 On the other hand, if   
  $x^e =1$  then the structure of  payoffs is  reversed: 
  $${\bf b_0} \quad \quad \begin {matrix} & {\bf a_0}  & {\bf a_{1} }  \cr 
{\bf a_0} & 0 & 1000 \cr 
 {\bf a_{1} } & 2000 & 0   \end{matrix} \quad \quad \quad \quad 
 {\bf b_1} \quad \quad \begin {matrix} & {\bf a_0}  & {\bf a_{1} }  \cr 
{\bf a_0} & 2000 & 0 \cr 
 {\bf a_{1} } & 0 & 1000   \end{matrix}$$
\vskip.1cm 
\noi A strategy of a player is a function from its collection 
 of  information sets to the   probability distributions on its 
 two actions (a one dimensional simplex).
 The strategy is Borel measurable if 
 that function is measurable with respect to its Borel sigma algebra (which 
 is defined canonically as above from its information sets).  
\vskip.2cm 

\noi Notice that however the $G_0$ player acts at some $(g,x)$, 
 that action is copied at $(r,x)$ (because the $G_0$ player cannot 
 distinguish between these two points). However the $R_i$ players 
 respond at $(r,x)$, those actions are copied at the sets 
 $\{ g\} \times T_i^{-1}(x)$ respectively (as the $R_i$ player 
 cannot distinguish between $(r,x)$ and  $\{ g\} \times T_i^{-1}(x)$). 
 The behaviour  of a player at  $(g,x)$ or $(r,x)$ will 
 influence inductively  the behaviour of all  players at an uncountable 
 subset leading upward through repetitive applications of the $T_i^{-1}$. 
However  
 the behaviour of  players that influences inductively 
 a player's payoff  at $(g,x)$ or $(r,x)$  
 lie entirely within the countable set $D\times G^+ (x)$.  
 With regard to this latter  aspect of influence,  
  our game shares similarity   with those defined by 
 countable equivalence relations.   

\section {No Harsanyi $ \frac 1  {1000}$ -equilibria} 
\noi Before we show that the game has no Harsanyi $\frac 1 {1000}$-equilibrium, 
 we focus in on the subset $\{ g\} \times X$.  
\vskip.5cm 
\noi Let $A_0$ be the subset of $ X$ such that the probability that  
 Player $G_0$ at $\{ g\} \times  A_0$ 
 chooses ${\bf b_0}$ is  at least $\frac {19} {20}$. 
 Let $A_1$ be the  corresponding 
 subset of $ X$ such that the probability that  
 Player $G_0$ chooses ${\bf b_1}$ is  at least $\frac {19} {20}$. 
 Let $A_M$ be the subset $ X \backslash (A_0 \cup A_1)$. 
\vskip.2cm \noi 

\noi As a general rule, from the above payoff matrices and the 
 assumption that players are following their interests (the interests  
 of the $R_i$ players at $(r,x)$ being that of conveying the choice 
 of the $G_0$ player at $(g,x)$),   
  we would expect that if $ T_1 (x)\in  A_i$ and 
 $ T_2 (x) \in  A_j$, and $x^e=k$ then 
 $ x\in  A_{i+j+k}$ where $i+j+k $ is represented modulo two. 
We call this the {\em parity rule}, and say that this  rule holds for 
 a point $(g, x)$  whenever 
  these three containments are true. We say that the parity rule 
 holds  for any $x\in X$ when 
 it holds for $(g, x)$.  
\vskip.2cm 
\noi 
If any player chooses both actions at some point 
 with strictly more than  $\frac 1{20}$ we say that the player is 
{\em mixing} at that point (meaning  
 in  $A_M$ when this player is $G_0$). If there is  a player and 
 a  set $A$ of measure at least 
 $w>0$ where  that player prefers one strategy over another 
 by at least  $r>0$ 
 and either that player is mixing or 
 choosing the non-preferred action, then that player can gain 
 at least $\frac {rw} {20}$ by choosing a different strategy. Therefore 
 in an $\epsilon$ equilibrium it follows that 
 $w$ is at most $\frac {20 \epsilon } r$. This simple fact is the 
 bridge between the equilibrium concept and the semi-group action on $X$.
\vskip.2cm 
\noi  With respect to an $\epsilon$-equilibrium for sufficiently small 
 enough $\epsilon$, 
there are two  aspects of the game very  important to our following arguments,  
 First, where the strategies in approximate equilibrium are not mixing, 
 they tend to fall into the parity rule and stay there. Second, 
 mixing is strongly discouraged by the structure of the payoffs.  
Looking at the payoffs of the $R_1$ and $R_2$ players at $(r, x)$,
 it is not possible for $G_0$ at  $(g,x)$ and $(r,x)$   
 to make both other players  indifferent to their two different actions. 
 And then if $z_0$ and $z_1$ are twins, namely 
 $T_i (z_0)= T_i (z_1)$ for both 
 $i=1,2$, if   the $R_i$ player is not mixing  
  at $(r, T_i(z_j))$ for at least one of $i=1,2$, 
 it is  not possible for the $G_0$ player at $(g, z_j)$ to 
 be indifferent to its two actions at both $j=0,1$.  This dynamic is 
 formalised in the next lemma. \vskip.2cm   
\noi {\bf Lemma 1:}   For every $x\in X$,   
 either one or the other  corresponding 
 Player $R_1$ or $R_2$ at  
 $(r, x)$  has an incentive of at least $80$ to choose 
 either ${\bf a_0}$ or ${\bf a_1}$ over the other action.  
 Let  $x,y\in X$ be any two points  in $X$ and let 
 $z_0$ and $z_1$ be the  
  two twins  where 
    $T_1 (z_i) = x$ and $T_2(z_i)=y$ for $i=0,1$ and $z_0^e=0$ and $z_1^e=1$.  
If one of $R_1$ or $R_2$  is mixing at $(r, x)$ or $(r, y)$,
 respectively, and 
 the other is not,   
 then    Player $G_0$ at either  $(g, z_0)$ or at $(g,z_1)$ 
 has an incentive of at least $80$ to choose either  ${\bf b_0}$ or 
 ${\bf b_1}$ over the other strategy.
\vskip.2cm 
\noi {\bf Proof:} Without loss of generality assume that $x^e=0$ and 
 that the  Player $G_0$ at $(g, x)$ chooses 
 ${\bf b_0}$ with probability at least $\frac 12$. By choosing ${\bf a_1}$ 
 the $R_1$ player would get no more than $50$ and by choosing 
 ${\bf a_0}$ the $R_1$ player would get at least $150$. On the 
 other hand, if the  Player $G_0$ at $(g, x)$ chooses 
 ${\bf b_1}$ with probability at least $\frac 12$ then the $R_2$ player 
 would get no more than $50$ by choosing ${\bf a_0}$  and at least 
 $150$ by choosing ${\bf a_1}$. \vskip.2cm 
\noi Next, due to symmetries, it suffices to consider the two cases 
 of the $R_2$ player choosing ${\bf a_0}$ with probability no more than 
 $\frac 1 {20}$ and the $R_2$ player choosing ${\bf a_1}$ with probability 
 no more than $\frac 1 {20}$. \vskip.2cm 
\noi Let $w\leq \frac 1 {20}$ be the probability that 
 the $R_2$ player chooses $\bf a_0$. 
 We break  this case into two subcases, where Player $R_1$ 
 chooses ${\bf a_0}$ with at least $\frac 3 5$ and where 
 Player $R_1$ chooses ${\bf a_0}$ with at most  $\frac 3 5$.  
 If Player $R_1$ chooses ${\bf a_0}$ with at least $\frac 35$ then the 
 $G_0$ player at $(g,z_1)$ gets at least $570$ for playing 
 ${\bf b_0}$ and no more than $400 (1-w) +2000w $ for playing 
 ${\bf b_1}$, which reaches a maximum 
 of $480$ at $w=\frac 1 {20}$.
 If Player $R_1$ chooses ${\bf a_1}$ with at least $\frac 25$
  then the $G_0$ player at $(g, z_0)$ gets at least $760$ from choosing 
 ${\bf b_0}$ and no more than 
 $600(1-w) + 2000w$ for playing ${\bf b_1}$, which reaches 
 a maximum of $670$ at $w= \frac 1 {20}$.  
 \vskip.2cm 
\noi  Now let  $w\leq \frac 1 {20}$ be the probability that 
 the $R_2$ player chooses $\bf a_1$. 
 We break  this   case into two subcases, where Player $R_1$ 
 chooses ${\bf a_0}$ with at least $\frac 3 5$ and where 
 Player $R_1$ chooses ${\bf a_0}$ with at most  $\frac 3 5$. 
     If Player $R_1$ chooses ${\bf a_0}$ with at least $\frac 3 5$ then 
 the $G_0$ player at $(g, z_1)$ get at least $1140$ from 
 choosing ${\bf b_1}$ and no more than 
 $820$ by choosing ${\bf b_0}$. If Player 
 $R_1$ chooses ${\bf a_1}$ with at least $\frac 25$ 
 then the $G_0$ player at $(g, z_0)$  gets at least $780$ from choosing 
 ${\bf b_1}$ and no more than 
 $600(1-w) + 2000 w$ from choosing ${\bf b_0}$, which reaches 
 a maximum of $670$ at $w=\frac 1 {20}$. \hfill $\Box$\vskip.2cm  
\noi The consequence of Lemma 1 is that the players are hardly ever mixing in an approximate 
 equilibrium. That is formalized in the next lemma. \vskip.2cm 
\noi {\bf Lemma 2:} In any $\frac 1 {1000}$ Borel measurable equilibrium 
 of the game, the $G_0$ player mixes with probability less 
 than $\frac {16} {10,000}$ and  the parity rule holds for all but at most 
$\frac 4 {1000}$ of  the space $X$.   
\vskip.2cm 
\noi {\bf Proof:}   
    Let $B_1$ be the subset 
 of $z\in A_M$ such that the corresponding $R_1$ player at 
 $(r, T_1 z)$ is mixing and let $B_2$ be the subset of 
 $z\in A_M$  such that the corresponding $R_2$ Player at 
 $(r, T_2 z)$ is mixing. Let $c= m(A_M)$, $a= m (B_1)$ and $b= m(B_2)$.    
  As  
 the $T_1 z$ and $T_2 z$ are distributed independently as variables 
 of $z$,  
 in an $\frac 1{1000}$ equilibrium the following holds: 
$c\leq \frac 1{4000} + a b + \frac12
 \big( a + b\big) $, where the $\frac 1 {4000}$ refers to the 
  maximum probability for the $G_0$ player 
  to choose a strategy that is suboptimal 
 by a quantity of at least $80$, the $ab$ refers to the probability 
 that both $R_i$  are mixing at both 
 $T_1 z$,  and 
 $\frac {a+b}2$ refers to the probability that Player $G_0$'s actions 
 are within $80$ of each other for one but not both of 
  the twins $z_0$ and $z_1$ (where one or the other of $R_1$ at 
 $T_1 z_i$ or $R_2$ at $T_2 z_i$ are mixing, but not both).    
  But as the sets $B_i$ are the sets $T^{-1}_i$  applied 
 to where $R_i$ is mixing in  $\{ r\} \times X$ and the $T_i$ are measure 
 preserving,  
   $a$ is also the probability throughout $\{ r\} \times 
 X$  that $R_1$ is mixing and  the same holds for $b$ and $\{ r\} \times 
 X$.   
 By Lemma 1, for any $z\in X$  the 
 probability of both $R_1$ mixing at $(r,z)$ and 
 $R_2$ mixing at $(r, z)$
  cannot exceed $\frac 1 {4000}$
 (from $\frac {20} { 80 \cdot 1000} =   \frac 1 {4000}$). 
   From this we conclude that  
 $a+ b \leq c + \frac 1 {1000}$, since  where $(g, z)$ is mixing at most 
 $\frac 1 {2000}$ of the points following in both  $(r, z)$ and 
 $(r, z)$ can be mixing, (with the other $\frac 1 {2000}$ referring 
 to the possibility that $G_0$ is not mixing at $(g, z)$ nevertheless 
 one of the $R_i$ players at $(r, z)$ is mixing). \vskip.2cm 
\noi From $ab\leq \frac 14 (a+b)^2$,   and the above,  we  
 get the quadratic  
  $0 \leq c^2 - \frac {999} {500}c + \frac {3,001} {1,000,000}$.
  After completing the square  we get  that 
 $|c-\frac {999}{1000}| \geq \sqrt {.995}$. Since $c$ cannot be greater than 
 $1$ we are left with 
 $c< .999-.9974 =.0016$.  
 The probability that the parity rule is not followed for an $x\in X$
 is no more 
 than  the probability of the $G_0$ player mixing at either 
 $(g, T_1 x)$ or $(g, T_2 x)$  plus the probability that 
 the $R_1$ player at $(r, T_1x)$, the $R_2$ player at $(r, T_2 x)$ or
 the $G_0$ player at $(g, x)$ is not properly responding to the 
 corresponding non-mixing behaviour. These probabilities sum 
 to  $.00395$. 
.    \hfill $\Box$
 \vskip.2cm 
\noi Next we show  it is impossible for there to exist 
 a $\frac 1 {1000}$ equilibrium Borel measurable equilibrium,  
 using the regularity of the measure. 
\vskip.2cm 
\noi 
  Let ${\cal C}_n$ be the set of cylinders of depth $n$, where the two 
 cylinders defined by the values $x^e=0$ and $x^e=1$ have depth $0$. 
 With $2^{n+1}-1$ words of length $n$ or less the cardinality of ${\cal C}_n$ 
 is $2^{2^{n+1}-1}$ and $m(c) =  2^{-2^{n+1}+1}$ for all $c\in {\cal C}_n$.   
    Recall the definition of $A_0$ and $A_1$ as the sets 
 where either $0$ or $1$ is played by $G_0$ with probability 
 at least $\frac {19} {20}$. For every 
 $c\in {\cal C}_n$ and $i=0,1$ let $w_i(c)$ be the conditional 
 probability of action ${\bf b_ i}$ at the cylinder $c$, in other 
words 
 $m(A_i \cap c) / m(c)$. For every cylinder
 $c$ define $\eta(c):= \min_{i=0,1} w_i(c)$ and let 
 $r(c)$ be the conditional probability of belonging in the set 
 where the parity rule does not hold. 
 \vskip.2cm 
\noi In the next lemma, we show that the parity rule is a powerful 
 force to equalize the probabilities for both actions ${\bf b_0}$ 
 and ${\bf b_1}$. This cannot be guaranteed for all cylinders, due
 to the small probability that the parity rule doesn't hold. But it 
 does hold in general for most  cylinders, regardless of the 
 depth. Two free generators and the dual causation implicit in 
 the parity rule   force this equalization.\vskip.2cm 

\noi {\bf Lemma 3:} In any $\frac 1 {1000}$ equilibrium of the game, 
  the average $q_i= \frac {\sum_{c\in {\cal C}_i} \eta(c)} {|{\cal C}_i|} $ is 
 at least $\frac 13$ for every $i$. \vskip.2cm  
\noi {\bf Proof:} The proof is by induction. There are 
 two elements in ${\cal C}_0$ and eight elements in ${\cal C}_1$. 
 Let $c_0$ and $c_1$  be  the two elements of ${\cal C}_0$.  
 Both cylinders 
 $c_0$ and $c_1$ are composed of four elements of ${\cal C}_1$, created (by means of $T_1$ and $T_2$) 
 from the combination of 
 the two elements $c_0, c_1$ of ${\cal C}_0$ and the 
 same two elements $c_0, c_1$  of ${\cal C}_0$ along with 
 a value of $x^e=0$ for all $x\in c_0$ and $x^e=1$ for all $x\in c_1$. 
 Let $x_0$ and $x_1$ be two points such that $x_0^e =0$, $x_1^e=1$, and 
 $x_0^U = x_1^U$ for every other $U\not= e$. 
However membership in $A_0$ or $A_1$   is determined by  
  $T_1 x_i$ and $T_2 x_i$, the parity rule requires  opposite memberships 
 for $i=0,1$. As the parity rule must hold in a set of size at least  
 $1-\frac 1{250}$, it follows that in the whole space the probability 
 given to both $A_0$ and $A_1$ must be approximately the same, more 
 precisely these probabilities must be at least 
  $\frac {124} {250}$ for both $A_0$ and   $A_1$.
 Now let $c$ be  either $c_0$ or $c_1$. 
 As $c$ is created by either the $e$ position being $0$ or $1$ and 
  the four combinations of $c_0$ and $c_1$ in 
 both direction  $T_1$ and $T_2$, whatever are the probabilities 
  given for the two $w_i(c_0)$ and the two $w_i(c_1)$, the fact that 
 $ \frac {w_i(c_0) + w_i(c_1)} 2 \geq   \frac {124} {250}$
 for both $i=0,1$, implies 
 that the conditional  probability given to both $A_0$ and $A_1$ at $c$ must be 
 at least $2(\frac {124} {250})^2 - \frac 1 {125}\geq .48$. 
   \vskip.2cm 
\noi 
We assume the claim is true for every $t\in {\cal C}_i$.   
 Every $t\in {\cal C}_{i+1}$ is created through the combination 
 of   a pair $c,d\in {\cal C}_i$ with a determination of $0$ or $1$ (though 
 this determination will play no rule in the following argument). 
 Let $i_c$ be the action that is less frequent at $c$, and 
 define $i_d$ the same way. Let $j$ be the action
 following from the parity rule determined by 
 the value of  $t^e$ and the combination of 
$i_c$ with $i_d$ (however that is determined by 
  the $e$ position).   If $r(c)= r(d)= r(t)=0$ then
  the parity rule would give $j$ exactly 
 $\eta(c) (1-\eta(d)) + \eta(d) (1-\eta(c))$, as it would give 
 the other action  
 the greater quantity 
 $(1-\eta(c)) (1-\eta(d)) + \eta(c) \eta(d)$.  
Due to the influence of the quantities $r(c), r(d), r(t)$ we cannot 
 say for sure that $j$ is the action less taken at $t$. But we can 
 say that  
 $\eta(t) \geq -r(t) + \eta(c) (1-\eta(d)-r(d)) + \eta(d) (1-\eta(c)-r(c))\geq 
 \eta(c) +\eta(d)- 2\eta(c) \eta(d)) -r(t) - \frac {r(c)+r(d)}2 $.    
  But with $\sum_{c\in {\cal C}_{j} } r(c) \leq \frac 1 {250} | {\cal C}_j|$
 for all $j$  
   it follows that  $ q_{i+1} \geq -\frac 1 {125} + 
  \frac 1 { |{\cal C}_{i}|^2} \sum_{c,d\in {\cal C}_{i} }  \eta(c) 
 +\eta(d)- 2\eta(c)\eta(d) =
  -\frac 1 {125} + \frac 1 {|{\cal C}_{i}|} \sum_{c\in {\cal C}_i} \eta (c) + 
  \frac 1 {|{\cal C}_{i}|}  \sum_{d\in {\cal C}_i} \eta (d) +  
 \frac 1  {|{\cal C}_{i}|^2} (\sum_{c\in {\cal C}_i} \eta (c)) ( \sum_{d\in {\cal C}_i} \eta (d) ) =  
 -\frac 1{125}+ 2 q_i - 2q_i^2$. By  induction 
 we conclude  that $q_{i+1} \geq \frac 49 - \frac 1 {125} > \frac 13$.   
\hfill $\Box$

 \vskip.2cm 
\noi {\bf Theorem  1:} There can be no Borel ( $\mu$)  measurable 
 $\frac 1 {1000}$-equilibrium. \vskip.2cm 
\noi {\bf Proof:} With $\eta(c)$ defined as in the proof of Lemma 3, 
for the  mutually exclusive measurable sets $A_0, A_1$ of $X$ it follows 
  from the regularity of the measure $\mu$ that $\lim_{n\rightarrow \infty} q_n = 
\lim_{n\rightarrow \infty} 
 \frac {\sum_{c\in {\cal C}_n} \eta (c)} {|{\cal C}_n|} =0$. But by Lemma 3 
  it never falls below $\frac 13$. \hfill $\Box$ 
\vskip.2cm 
\noi Notice that where  the players are 
 obeying the parity rule, even approximately so, the location where 
   the payoff to Player 
 $G_0$ is close to $2000$ or to $1000$ is determined by one or the other 
 of  the two other players, by Player $R_1$ 
 in the half $\{ x\in X \ | \ x^e=0\}$ and by Player $R_2$ in 
 the half   $\{ x\in X \ | \ x^e=1\}$. Where 
 the parity rule holds  the Lesbegue 
measurability of the payoff of Player $G_0$ 
 implies the Lesbegue measurability of the strategies 
 of the $R_i$ players and hence also the Lesbegue measurability of 
 the strategy   of the $G_0$ player in response.  For the $G_0$ player  to 
 have an ``equilibrium payoff'' by some interpretation one must define 
 that concept quite differently   from the existence of Harsanyi 
$\epsilon$-equilibria.

\section {Bayesian equilibria:} 

In this section we will show that we can colour
 the space $X=\{0,1\}^{G^{+}}$ modulo a null set $N$ using only
 two colours: $1$ and $0$ or red and blue, respectively, so that 
 the parity rule is obeyed, and 
 furthermore extend this to an equilibrium of the whole space $\Omega$. 
   Recall that the 
\emph{parity rule} is a function $c: X \rightarrow X$ 
such that $c(x) = c(T_1(x)) + c(T_2(x)) + x^e$ (modulo 2). Notice 
 that such a colouring on such a set defines a Bayesian equilibrium on 
 the subset $\{ g, r \} \times (X\backslash N)$. 
 We show then how to extend these strategies to a Bayesian equilibrium 
 on all  of the space  $\Omega = \{ g,  r\} \times X$.
\\
 Recall the definition 
 of the twins determined by some $x,y\in X$.  
 We say that 
 a subset $A$ of $X$ is closed if for every pair $x,y$ in $A$ 
 the twins determined by $x$ and $y$ are also in $A$. 
 By the \emph{closure} $\overline A$  of a set $A \subseteq X$ we mean
 the smallest closed set
 containing $A$. 
 We say that $A \subseteq X$ is \emph{pyramidic} if $x \in A$ implies 
that  $ U(x) \in A$, where $U \in G^{+}$.
 The main example of pyramidic set is  $G^{+}(x)$,
 where $x$ is an arbitrary element of $X$. 
 Notice that whenever a set is pyramidic that its closure  is also
 pyramidic.\\ 

\noi 
\indent Define now the set
  $$ N = \{x \in X: U(x) = V(x), \mbox
 { for some distinct } U, V \in G^{+} \}.$$
 This set is a null set with respect to the product measure $m$ on $X$.
 Indeed, for any given two distinct 
  words $U, V \in G^{+}$, 
the equality implies an agreement on infinitely many coordinates,
 and there are only  countably many words $U \in G^{+}$.\\   
 \noi  We are ready to prove the following lemma:\\

\vskip.2cm 
  \noi {\bf Lemma  4:} Let $X_1 = X \setminus N$
 and let $c$ be the parity rule defined on $X$.
 Then there exists a colouring of $X_1$ using the 
 two colours $\{ 0,1\}$ which is consistent
 with the rule $c$.\\

\noi  {\bf Proof:} We will proceed by transfinite induction, 
 assuming the Axiom of Choice and thus 
 Zorn's Lemma. 
Let $x_0$ be any element of $X_1$
 and obtain the set 
$P_0 :=G^{+}(x_0)$. We define now the colouring of  $P_0$
 of $G^{+}(x_0)$ as follows:\vskip.2cm 
  
  \noi  $(i)$ colour all the points $T_1U(x_0)$ in red,
 where $U \in G^{+}(x_0)$ and $U$ is the identity or begins on the
 right with $T_1$;     \vskip.2cm 
  \noi  $(ii)$ colour all the points $T_2V(x_0)$ in blue,
 where $V \in G^{+}(x_0)$ is the identity or begins on the right with $T_1$; 
  \vskip.2cm 
\noi  $(iii)$ colour the remaining points of the pyramid $P_0$ 
  in the way that they satisfy the rule $c$.
\vskip.2cm 
\noi After colouring all the points of $P_0$, extend the colouring 
 to all the points in the closure $\overline {P_0}$ of $P_0$.    
 \vskip.2cm 

 \noi  Next create a partial ordering on colourings 
 of pyramidic and closed subsets 
 of $X$ that obey the parity rule, with one colouring greater than another 
 if  the subset is larger and their colourings agree on their
 common intersection (the smaller subset). Any tower of such colourings 
 will define  a colouring that obeys the parity rule.   As Zorn's Lemma 
 implies that there is a maximal element (as a tower defines  its own least 
 upper bound), it suffices to show that 
 maximality implies that all of $X_1$ has been coloured.  Let $P$ be 
 any pyramidic and closed subset of $X_1$ with a colouring 
 that obeys the parity rule. Assume that $P$ is not all of 
 $X_1$.  Let $x$ be the first member of $X_1$ 
 that is not in $P$. 
\vskip.2cm  
 \noi We say
 that $x$ has a \emph{hitting point in } $P$ if
 $U(x) \in P$ for some
 $U \in G^{+}$ and whenever 
  $U= VW$ and $W$ is not the identity then 
 $V(x) \not\in P$.     \vskip.2cm 
  
\noi Now 
   we have the two cases:\\

  \noi Case 1) $x$ has no hitting point with respect to $P$.
 Then we colour the closure $ G^+ (x)$ in the same way as  
the initial pyramidic set $P_0$.\\

  \noi Case 2) $x$ has a hitting point in  $P$. 
  Then colour the elements of
 $G^{+}(x)$ taking into an account the colours of the hitting points. 
  Notice that the closure of $P$ implies that if $Ux$ is not in $P$ 
 then one of $T_i U x$ is also not in $P$, so that if $T_i U x$ is a hitting 
point then $T_j Ux$ is not a hitting point, for $i\not=j$. 
 This allows us to colour $x$ arbitrarily and then 
 move downward in a consistent way, with the colouring 
 of $T_iUx$ determined already only if $T_j Ux$ is a hitting point 
 or had just  been just coloured arbitrarily (for $j\not=i$). \\

  \noi And then we colour the closure of $G^+ (x) \cup P$ 
 according to the parity rule $c$ for a larger 
 set that is closed, pyramidic, and consistent with both  the parity 
 rule and the pre-existing colouring. \hfill    $\Box$\\

\noi {\bf Theorem 2:} There exists a Bayesian equilibrium on
 all of $\Omega$.\\

\noi {\bf Proof:} Following on from the proof of Lemma 4, let 
  $\overline {X_1}$ be the closure of $X_1$ and extend the colouring 
 of $X_1$ to one on $\overline {X_1}$.   For all three 
 players $G_0$, $R_1$ and $R_2$ define pure strategies on 
 $D \times \overline {X_1}$ accordingly. 
 With $x$ any point in $X\backslash X_1$, let 
 $\Gamma _x$ be the game defined on 
  $D \times G^+ (x)$ such  that 
the strategies on  $D \times (G^+ (x) \cap \overline {X_1})$
 are 
 already fixed  by the above colouring.
 As the game has only countably many positions, by Simon (2003) there 
 is a Nash equilibrium defined on the game $\Gamma_x$. But notice 
 that it defines an equilibrium when including 
 those strategies on  
 $D \times (G^+ (x) \cap \overline {X_1})$
that  are 
 already fixed.  
 Extend this equilibrium to an equilibrium on the set 
 $D \times (\overline {G^+ (x) \cap \overline {X_1}})$ through 
 best reply responses (noticing that nothing done at a point $y$ has 
 any influence on any player at points $Uy$ for any $U\not= e$ - just consider $x, T_1x, T_2x$ and follow by induction).  
We can even assume that these best reply responses are pure strategies. 
 As with the proof of Lemma 4, we define a partial ordering on pyramidic 
 and closed  subsets 
 $P$ of $X$ and the  equilibria defined on 
 $D \times P$. In the same way, we show that an equilibrium 
 can be defined on all of $\Omega$, using critically that any extension 
 of an equilibrium from  a closed and pyramidic set $P$ 
 will not disturb the pre-existing equilibrium property on $P$.  
\hfill  $\Box$ 
\vskip.2cm 
 \noi There are some points in $X$ for which any equilibrium 
  requires a mixed strategy. 
 Let $x,y$ be the two points defined by $x^e=0$, $y^e=1$, $T_1 (x)= y$, 
 $T_2 (x) =x$, $T_1 (y) =x$ and $T_2 (y) =x$. No matter how 
 $x$ is coloured, because $T_1 (y) = T_2 (y)=x$ and $y^e=1$, 
  $y$ must be coloured with $1$.   But then  $T_1 (x)= y$, 
 $T_2 (x) =x$ and $x^e=0$ forces $x$ to be coloured differently from itself. 

\section {Conclusion: open questions} 

\noi Is there an example of an ergodic game (Simon 2003) that has 
 no Harsanyi $\epsilon$-equilibrium for some positive $\epsilon$? 
The examples of Simon (2003) and  of Hellman (2014) were ergodic games, 
 and ergodic games have Bayesian equilibria (Simon 2003). We believe 
 the answer is yes and that it can be done through the action of a 
 non-amenable group which defines 
 the information structure of the players.
  With  the example of this paper, there was 
 a very strong mixing structure that kept the probability high 
 for both actions at all cylinders.   
  We believe that the  weaker mixing structure from a group 
 action would be sufficient to obtain the same result. \vskip.2cm 
\noi In the example of this paper, there are three players. Can the 
 same result be accomplished with two players? We believe that 
 the answer is yes.  The  structure of our example is similar to 
 that of 
   a free  action of the free product $G=C_2 * C_3$ on $\{ 0,1\} ^G$. 
 We believe it can be done through  associating  $C_3$ with the knowledge 
 of  one player and 
 $C_2$ with  the other player, (or with   two other finite groups 
 whose free product is non-amenable).  \vskip.2cm 
\noi Lastly, what is the relationship between Bayesian equilibria and  the 
 Banach-Tarski paradox? Let $G$ be a group  
 acting in a measure preserving way on a probability
 space $X$ and for every player $i$ let 
 $G_i$ be a finite subgroup of  $G$  so that  
 the information sets of Player $i$ are the orbits of $G_i$ and $G$ is 
 generated by the $G_i$. 
 Is there a 
 Bayesian game so defined 
 such that every Bayesian   
 equilibrium  defines through its 
 fibres  a paradoxical decomposition of the space where the   
 group elements  demonstrating the paradox belong to $G$?  On this 
 question we are agnostic.

\section  {References}

\begin{description}

\item[Harsanyi, J. (1967)], Games with Incomplete Information Played 
 by Bayesian Players, {\em Management Science}, Vol. 14, pp. 159-182.

\item [Hellman, Z. (2014)], A Game with No Bayesian Approximate 
 Equilibria, {\em Journal of Economic Theory}, Vol. 153, pp. 138-151. 

 \item [Hellman, Z. and Levy, J. (2016)], Bayesian Games over Continuum Many 
 States, 
{\em Theoretical Economics}, forthcoming. 

\item[Kechris, A., Solecki, S. Todorcevic, S. (1999)], Borel Chromatic 
 Numbers, Vol 141, Issue 1, pp. 1-44. 

\item [Milgrom, P. and Weber, R. (1985)], Distribution 
 Strategies for Games with Incomplete Information, {\em Mathematics 
 of Operations Research}, Vol. 10, No. 4, pp. 619-632. 

\item[Simon, R. S. (2003)] , Games of Incomplete Information, Ergodic 
Theory, and the Measurability of Equilibria,
 {\em Israel J. Math.}, Vol. 138, No. 1, pp. 73-92.

\end{description}

\end{document}